\documentclass[aps,prc,twocolumn,superscriptaddress]{revtex4-2}
\usepackage{graphicx}
\usepackage{makecell}
\usepackage{color}
\usepackage[dvipsnames]{xcolor}
\usepackage[colorlinks=true,urlcolor=Blue,linkcolor=Blue]{hyperref}
\usepackage[all]{hypcap}
\usepackage{amsmath}
\usepackage{multirow}
\usepackage{booktabs}
\usepackage{soul}
\usepackage{lineno}

\begin{document}

\title{Effective proton-neutron interaction in mirror nuclei}

\author{Y.~M.~Xing}
\affiliation{CAS Key Laboratory of High Precision Nuclear Spectroscopy,   Institute of Modern Physics, Chinese Academy of Sciences, Lanzhou 730000, China}
\affiliation{School of Nuclear Science and Technology, University of Chinese Academy of Sciences, Beijing 100049, China}
\author{Y.~F.~Luo}
\affiliation{CAS Key Laboratory of High Precision Nuclear Spectroscopy,   Institute of Modern Physics, Chinese Academy of Sciences, Lanzhou 730000, China}
\affiliation{School of Nuclear Science and Technology, University of Chinese Academy of Sciences, Beijing 100049, China}

\author{K. H. Li}
\affiliation{Institute of Particle and Nuclear Physics, Henan Normal University, Xinxiang 453007, China}
\affiliation{CAS Key Laboratory of High Precision Nuclear Spectroscopy,   Institute of Modern Physics, Chinese Academy of Sciences, Lanzhou 730000, China}

\author{Y.~H.~Zhang}
\email{yhzhang@impcas.ac.cn}
\affiliation{CAS Key Laboratory of High Precision Nuclear Spectroscopy,   Institute of Modern Physics, Chinese Academy of Sciences, Lanzhou 730000, China}
\affiliation{School of Nuclear Science and Technology, University of Chinese Academy of Sciences, Beijing 100049, China}	

\author{J.~G.~Li}
\email{jianguo\_li@impcas.ac.cn}
\affiliation{CAS Key Laboratory of High Precision Nuclear Spectroscopy,   Institute of Modern Physics, Chinese Academy of Sciences, Lanzhou 730000, China}
\affiliation{School of Nuclear Science and Technology, University of Chinese Academy of Sciences, Beijing 100049, China}

\author{M.~Wang}
\affiliation{CAS Key Laboratory of High Precision Nuclear Spectroscopy,   Institute of Modern Physics, Chinese Academy of Sciences, Lanzhou 730000, China}
\affiliation{School of Nuclear Science and Technology, University of Chinese Academy of Sciences, Beijing 100049, China}

\author{Yu.~A.~Litvinov}
\affiliation{CAS Key Laboratory of High Precision Nuclear Spectroscopy,   Institute of Modern Physics, Chinese Academy of Sciences, Lanzhou 730000, China}
\affiliation{GSI Helmholtzzentrum f{\"u}r Schwerionenforschung, Planckstra{\ss}e 1, 64291 Darmstadt, Germany}

\author{K.~Blaum}
\affiliation{Max-Planck-Institut f\"{u}r Kernphysik, Saupfercheckweg 1, 69117 Heidelberg, Germany}	

\author{X.~L.~Yan}
\affiliation{CAS Key Laboratory of High Precision Nuclear Spectroscopy,   Institute of Modern Physics, Chinese Academy of Sciences, Lanzhou 730000, China}
\affiliation{School of Nuclear Science and Technology, University of Chinese Academy of Sciences, Beijing 100049, China}

\author{T.~Liao}
\affiliation{CAS Key Laboratory of High Precision Nuclear Spectroscopy,   Institute of Modern Physics, Chinese Academy of Sciences, Lanzhou 730000, China}
\affiliation{School of Nuclear Science and Technology, University of Chinese Academy of Sciences, Beijing 100049, China}

\author{M.~Zhang}
\affiliation{CAS Key Laboratory of High Precision Nuclear Spectroscopy,   Institute of Modern Physics, Chinese Academy of Sciences, Lanzhou 730000, China}
\affiliation{School of Nuclear Science and Technology, University of Chinese Academy of Sciences, Beijing 100049, China}

\author{X.~Zhou}
\affiliation{CAS Key Laboratory of High Precision Nuclear Spectroscopy,   Institute of Modern Physics, Chinese Academy of Sciences, Lanzhou 730000, China}
\affiliation{School of Nuclear Science and Technology, University of Chinese Academy of Sciences, Beijing 100049, China}

\begin{abstract}
Effective proton-neutron interactions, $V_{pn}$, in mirror nuclei are systematically analyzed using the ground-state atomic masses.
A mirror symmetry of $V_{pn}$ is found for bound nuclei with a standard deviation
of $\sigma=32$ keV. 
However, this mirror symmetry is apparently broken for some mirror-nuclei pairs  
when a proton-unbound nucleus is involved in extracting the $V_{pn}$ values. 
Such a mirror-symmetry breaking is attributed to the Thomas-Ehrman shift of the  proton-unbound nucleus and investigated by using the Gamow shell model. It is concluded that the Thomas-Ehrman shift originates mainly from reduced Coulomb energies in the proton-unbound nuclei due to the extended radial density distribution of valence protons. 
\end{abstract}
\maketitle

\section{Introduction}
Isospin symmetry was introduced under the assumption of charge symmetry and charge independence of the nuclear force~\cite{Wigner1937}. 
It is a useful concept in nuclear and particle physics, yielding startling symmetries in the behavior of nuclei near the $N = Z$ line. 
In reality, isospin symmetry is slightly broken due to the Coulomb interaction among protons, the proton and neutron mass difference, and the presence of the charge-dependence of the nuclear force. Examining and exploring the isospin-related symmetries, and determining the extent to which they are broken, has become an intriguing topic in nuclear structure physics. 

There are many probes to investigate isospin symmetry and its breaking, such as 
the mirror energy differences~\cite{Bentley2015,Bentley2022,Li2023}, Coulomb energy differences~\cite{Lenzi2001,Bentley2007}, and triplet energy differences~\cite{Henry2015,Lenzi2018,Bentley2022}. 
The proton-neutron ($p$-$n$) interaction was also used to explore the symmetry breaking of charge-dependent nuclear effects~\cite{Basu1971,PhysRevC.6.467}. 
Based on the concept of isospin symmetry, the wave functions of isospin analogue states are essentially identical, and the effective $p$-$n$ interactions are expected to be same in a pair of well-bound mirror nuclei. However, as one proceeds towards the proton dripline region, nuclear states become weakly bound or unbound against particle emissions, and the well-known Thomas-Ehrman shift (TES)~\cite{Thomas1952,Ehrman1951} will be more significant. Consequently, the expected identical $p$-$n$ interactions between mirror nuclei will be affected. 

Indeed, the $p$-$n$ interactions in the mirror nuclei, or more generally, in the isospin multiplets, were investigated over half a century ago~\cite{Basu1971,PhysRevC.6.467} although  available binding energies were limited at that time. Recently, developments of experimental technique 
have led to a wealth of new and accurate mass values of proton-rich nuclei (See Refs.~\cite{ZhangM2023,PhysRevLett.130.192501} for example), making it feasible to systematically examine the effective $p$-$n$ interactions in the mirror nuclei. 
The $p$-$n$ interactions in mirror nuclei were observed to be approximately equal~\cite{Zhang2018,ZhangM2023}. This property was utilized to make mass predictions for proton-rich nuclei~\cite{Zong2019}. 
Nevertheless, comprehensive study of $p$-$n$ interactions in mirror nuclei and investigation of mirror symmetry and its breaking are still lacking. 

In this work, we focus on the so-called empirical residual $p$-$n$ interactions, $V_{pn}$, of mirror nuclei extracted from the latest Atomic-Mass Evaluation AME20~\cite{AME2020} and the updated atomic masses~\cite{ZhangM2023,PhysRevLett.130.192501,Yu2024,Yandow2023,Campbell2024}. 
In Sec.~\ref{Comparison}, we demonstrate a mirror symmetry of ${V_{pn}}'$s if they are deduced from bound nuclei. We also show that such a mirror symmetry can be explained simply by using a charge-independent mass formula. 
In Sec.~\ref{Breaking}, we highlight the mirror-symmetry breaking of $V_{pn}$ for some mirror partners
when a proton-unbound nucleus is used in extracting the $V_{pn}$ value. 
In Sec.~\ref{GSM}, we conduct Gamow shell model calculation for a better understanding of such a mirror-symmetry breaking. 
A summary is given in Sec.~\ref{Summary}. 

\section{Mirror symmetry of $V_{pn}$ for bound nuclei}\label{Comparison}


Conventionally, one extracts the empirical residual $p$-$n$ interaction (or effective $p$-$n$ interaction), $V_{pn}$, from the binding energies, $B(Z,N)$, of four neighboring nuclei according to~\cite{PhysRevLett.74.4607}
\begin{equation}\label{eq:dv1}
	\begin{aligned}
		  V_{pn}^{ee}(Z,N)=&\frac{1}{4}[B(Z,N)-B(Z,N-2) \\
		&-B(Z-2,N)+B(Z-2,N-2)],
	\end{aligned}
\end{equation} 
\begin{equation}\label{eq:dv2}
	\begin{aligned}
		V_{pn}^{oe}(Z,N)=&\frac{1}{2}[B(Z,N)-B(Z,N-2) \\
		&-B(Z-1,N)+B(Z-1,N-2)],
	\end{aligned}
\end{equation}
\begin{equation}\label{eq:dv3}
	\begin{aligned}
		V_{pn}^{eo}(Z,N)=&\frac{1}{2}[B(Z,N)-B(Z,N-1) \\
		&-B(Z-2,N)+B(Z-2,N-1)],
	\end{aligned}
\end{equation}
\begin{equation}\label{eq:dv4}
	\begin{aligned}
		V_{pn}^{oo}(Z,N)=&[B(Z,N)-B(Z,N-1)- \\
		&B(Z-1,N)+B(Z-1,N-1)].
	\end{aligned}
\end{equation}

The extracted $V_{pn}$ values are considered to be a measure of effective $p$-$n$ interactions between a valence proton and a valence neutron in the nucleus with $Z$ protons and $N$ neutrons.  
Eqs.~(\ref{eq:dv1}) through (\ref{eq:dv4}) are applicable to nuclei with $(Z,N)$ configurations of odd-odd(oo), even-odd(eo), odd-even(oe), and even-even(ee) parities, respectively.  

 The above-mentioned definition of $V_{pn}$ is slightly different from that used in Refs.~\cite{Basu1971,PhysRevC.6.467}, yielding a more smooth behavior with mass number $A$~\cite{Wu2016}.
$V_{pn}$ plays a vital role in various nuclear structural phenomena, such as the onset of collectivity~\cite{Casten1988,Cakirli2006},  
changes of shell structure~\cite{PhysRevLett.94.092501}, and phase transitions in nuclei~\cite{HEYDE1985303,FEDERMAN1977385,FEDERMAN197829}.
Moreover, it served also as a criterion for theoretical investigations~\cite{Stoitsov2007,Wu2016}, and some regularities of $V_{pn}$ throughout the whole nuclear chart were used for mass predictions~\cite{Stoitsov2007,Fu2010}. 

We have extracted the $V_{pn}$ values via Eqs.~(\ref{eq:dv1}) through (\ref{eq:dv4}) using the experimental ground-state masses from the latest Atomic-Mass Evaluation AME20~\cite{AME2020}, and the recent results from the $B\rho$-IMS~\cite{ZhangM2023,PhysRevLett.130.192501,Yu2024} and the LEBIT mass measurements ~\cite{Yandow2023,Campbell2024}. 
The mass values of $^{29}$Cl, $^{30}$Ar, obtained from the recently measured $S_{p}$($^{29}$Cl)~\cite{Xu2018} 
and $S_{2p}$($^{30}$Ar)~\cite{Xu2018}, are also used in the analysis.
We compare the difference of ${V_{pn}}'$s between mirror nuclei according to the parameter defined as~\cite{ZhangM2023} 
  \begin{equation}\label{Eq:DdVpn}
  \Delta V_{pn}(T_z^{<},A)=V_{pn}(T_z^{<},A)- V_{pn}(T_z^{>},A),
 \end{equation}
where  $T_z^{<}$ and $T_z^{>}$ represent the negative and positive values of the isospin projection $T_z=(N-Z)/2$ for a mirror partner. 
The deduced $\Delta V_{pn}$ value
is then assigned to the nucleus with negative $T_z$.

\begin{figure}[htb]
	\centering
	\includegraphics[scale=0.33]{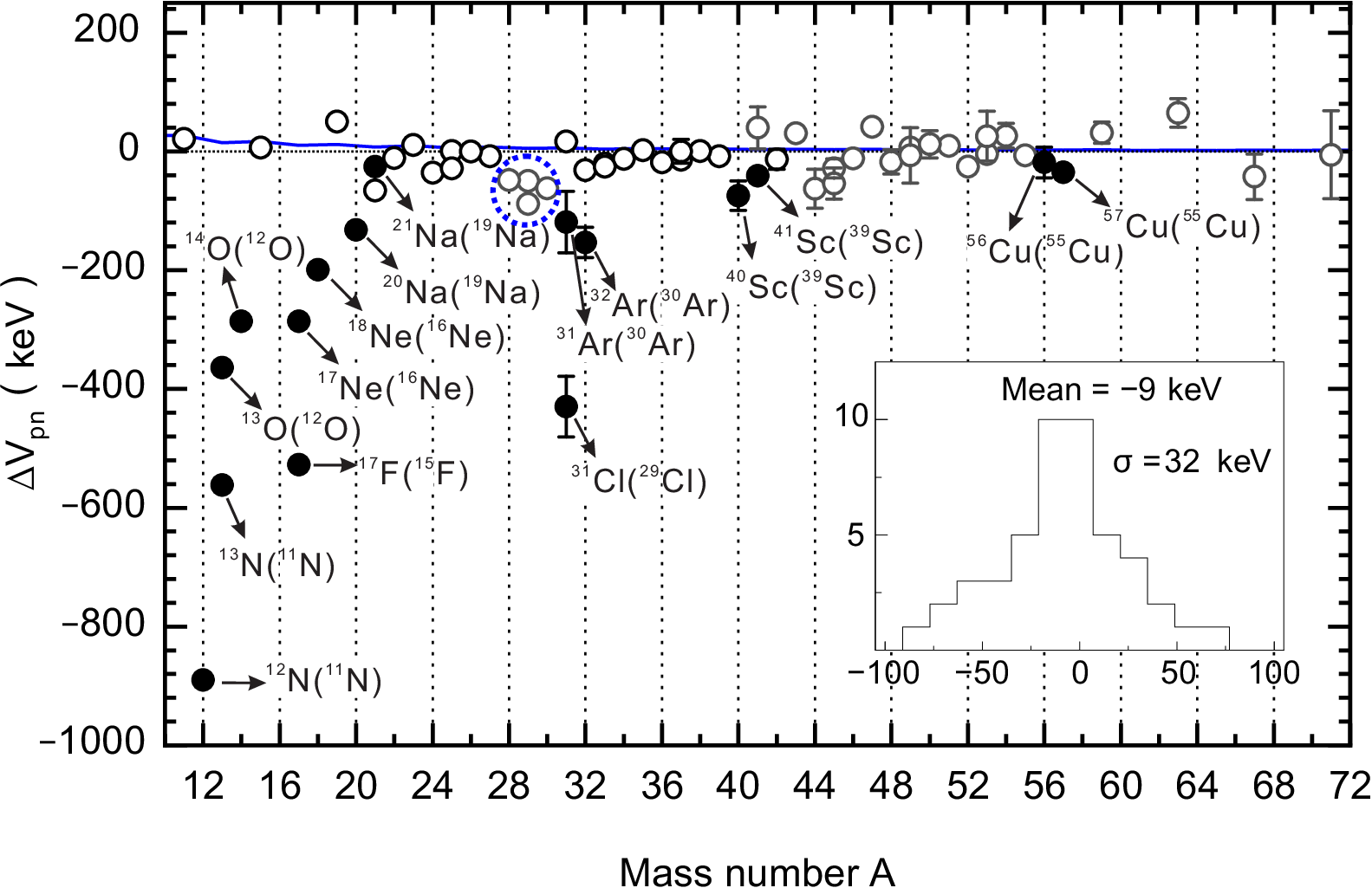}
	\caption{Plot of $\Delta V_{pn}$ values extracted using all bound nuclei (open circles) or those involving a particle-unbound nucleus (filled circles) marked in the brackets. The blue solid line represents the calculated $\Delta V_{pn}$ values using Eq.~(\ref{DdVpn_formular}) for $T_z=-3/2$ nuclei. The inset gives the distribution of $\Delta V_{pn}$ values extracted from the bound nuclei, showing a mirror symmetry of $V_{pn}$ at a confidence level of $\sigma=32$ keV. 
    Note that the four data points enclosed by the dotted blue circle represent ${\Delta V_{pn}}'$s of $^{28,29}$P and $^{29,30}$S, which are related to the proton halo nuclei ($^{27,28}$P and $^{28}$S). 
}
	\label{fig:DdVpn}
\end{figure}

As shown in Eqs.~(\ref{eq:dv1}) through (\ref{eq:dv4}), four nuclear masses are used in extracting a $V_{pn}$ value. 
In our analysis, we first use the bound nuclei to extract the ${\Delta V_{pn}}'$s. The results are presented as the open circles in Fig.~\ref{fig:DdVpn}. The inset shows the $\Delta V_{pn}$ distribution which is centered at the mean value of $-9$ keV with the standard deviation of $\sigma=32$ keV. This fact indicates that the mirror symmetry of $V_{pn}$ holds well at a confidence level of 32 keV when dealing with bound nuclei in the analysis.

It is worth noting that the $\Delta V_{pn}$ values of $^{28,29}$P and $^{29,30}$S are obviously negative, see the data points enclosed in the dotted blue circle in Fig.~\ref{fig:DdVpn}. This is most likely due to the proton-halo structure of $^{27,28}$P and $^{28}$S~\cite{Yu2024,Navin1998,Chu2008}. If these four data points are rejected, the $\Delta V_{pn}$ distribution will be centered at $-4$ keV with the standard deviation of $\sigma= 28$ keV.

The above-mentioned mirror symmetry of $V_{pn}$ can be understood simply by using the mass formula given in Refs.~\cite{Janecke1966,MacCormick2014}:
  \begin{equation}\label{M}
    M(A,T,T_z)=M_0(A,T)+E_c(A,T,T_z)+T_z\Delta_{nH},
 \end{equation}
where $M_0(A,T)$ represents the charge-free nuclear
mass, $E_c(A,T,T_z)$ the total charge-dependent energy in the nucleus (conventionally named as the Coulomb energy), and $\Delta_{nH}$ the neutron-$^1$H mass difference. 

Assuming that $E_c(A,T,T_z)$ is dominated by the Coulomb energy, and the nucleus is a uniformly charged sphere, one has a simple form of  
  \begin{equation}\label{Ec}
    E_c=a\frac{Z^2}{A^{1/3}}
 \end{equation}
with $a$ being a constant. Combining Eqs.~(\ref{eq:dv1}) though~(\ref{Ec}), one has an approximated expression 
  \begin{equation}\label{DdVpn_formular}
  \Delta V_{pn}\simeq\frac{4a|\overline{T_z}|}{9\overline{A}^{4/3}},
 \end{equation}
where $\overline{T_z}$ and $\overline{A}$ represent, respectively, the average  $T_z$ and $A$ of the four nuclei used in extracting the $V_{pn}$ of the proton-rich (or neutron-rich) nucleus.  
For $T_z=-3/2$ nuclei, the calculated $\Delta V_{pn}$ values 
with $a=0.69$ MeV are given in Fig.~\ref{fig:DdVpn} as the blue solid line. This line is 
in the proximity of zero, particularly in the heavier nuclear region, confirming the mirror symmetry of $V_{pn}$.  

\begin{figure}[htb]
	\centering
\includegraphics[scale=0.28]{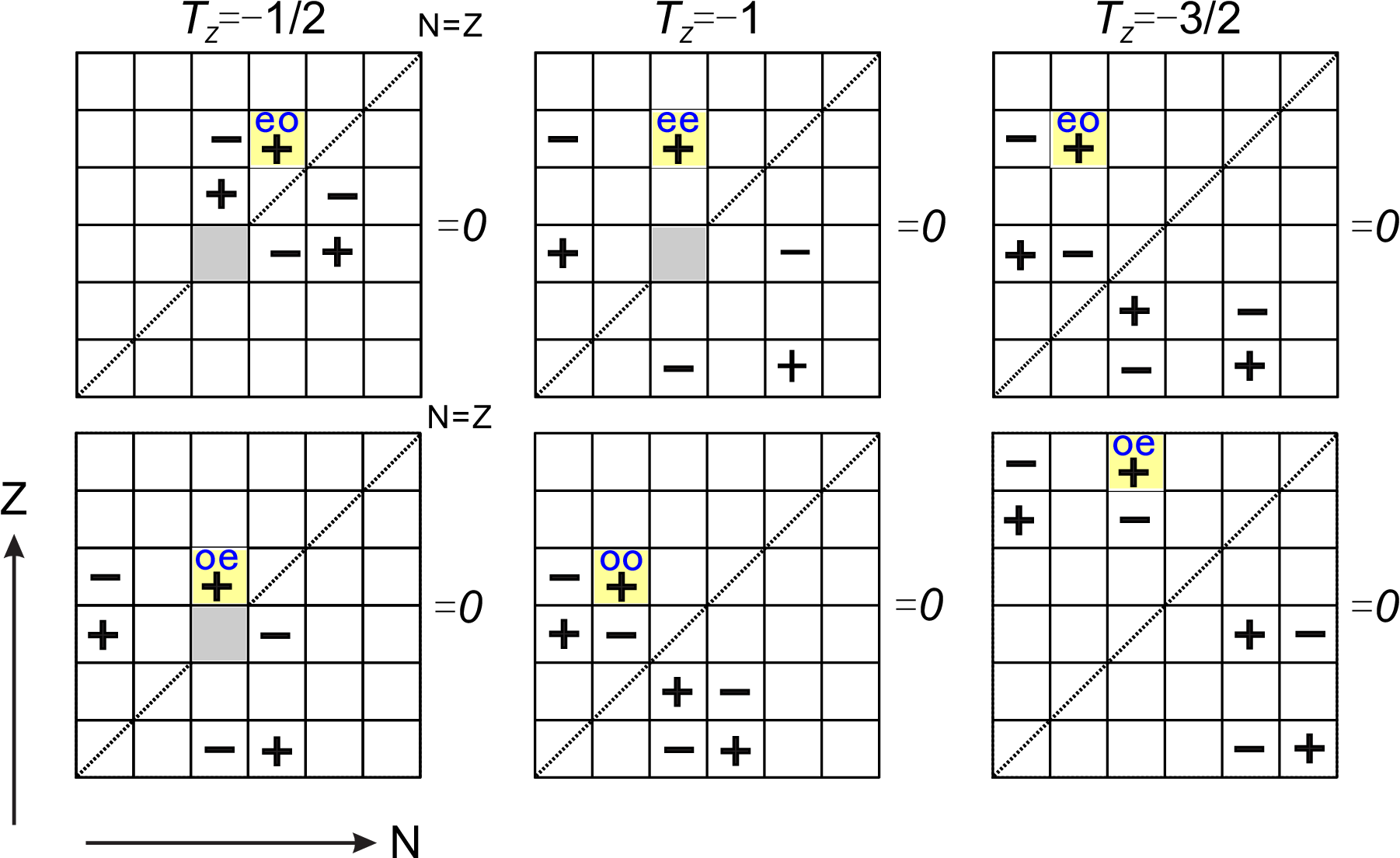}
	\caption{Schematic illustration of local mass relations based on the mirror symmetry of $V_{pn}$, i.e., the sum of the binding energies with plus or minus signs marked in the boxes equals to zero. The odd-even parity of the nucleus is highlighted in yellow. 
 The gray box indicates that the binding energy of the nucleus is used twice, and its contribution is canceled out. 
 }
	\label{fig:Chart_DdVpn}
\end{figure}

If the mirror symmetry of $V_{pn}$, i.e., $\Delta V_{pn}\simeq 0$ is adopted, a set of local mass relations can be obtained subsequently. 
Examples are depicted in Fig.~\ref{fig:Chart_DdVpn} where the $\Delta V_{pn}$ value is assigned to the nuclei (marked in yellow) with $T_z=-1/2, -1$, and $-3/2$, respectively. 
Such local mass relations are similar to those suggested in Refs.~\cite{Kelson1966,Bao2013,Tian2013}.  
Especially, by an appropriate combination of these simple local mass relations, the charge-symmetric Kelson-Garvey mass formula (see Eq.~(3) in Ref.~\cite{Kelson1966} or Fig.~6 in Ref.~\cite{Janecke1988}) and also the improved one in Ref.~\cite{Tian2013} can be obtained. 
This indicates that the mirror symmetry of $V_{pn}$ is a prerequisite of the charge-symmetric Kelson-Garvey relation. In other words, the charge-symmetric Kelson-Garvey relation originates in fact from the mirror symmetry of $V_{pn}$.   

\section{Mirror-symmetry breaking of $V_{pn}$}\label{Breaking}

In the previous section, we have shown that a mirror symmetry of $V_{pn}$ exists for bound nuclei at a confidence level of $\sigma=32$ keV. However, when adding more $\Delta V_{pn}$ data involving a particle-unbound nucleus in Fig.~\ref{fig:DdVpn} (see the filled circles), the so-called mirror symmetry of $V_{pn}$ is apparently broken for a few mirror pairs, 
i.e., the $V_{pn}$ values of the proton-rich nuclei are significantly smaller than those of the corresponding neutron-rich. The negative $\Delta V_{pn}$ values are probably caused by the fact that a proton-unbound nucleus is used in extracting the $V_{pn}$ value of the corresponding proton-rich nucleus. 
These $\Delta V_{pn}$ values are shown as the filled circles in Fig.~\ref{fig:DdVpn} and the proton-unbound nuclei are indicated in the parentheses. 

The observed mirror-symmetry breaking of $V_{pn}$ was attributed~\cite{PhysRevC.6.467} 
to the TES of the proton-unbound nucleus.
Taking $^{17}$F as an example, its effective $p$-$n$ interaction is calculated by
\begin{equation}\label{eq:dv5}
V_{pn}(^{17}{\rm F)}=\frac{1}{2}[B(^{17}{\rm F})-B(^{15}{\rm F})-B(^{16}{\rm O})+B(^{14}{\rm O})].
\end{equation}
Among the four nuclei, $^{15}$F is proton-unbound with the proton separation energy of $S_p=-1270(14)$ 
keV~\cite{AME2020}. The valance proton in $^{15}$F occupies predominantly the unbound $\pi s_{1/2}$ single particle orbit, and the wave function or density distribution would be spatially expanded, leading to a reduction of the Coulomb energy~\cite{PhysRevC.100.064303}. 
Since the Coulomb interaction is repulsive, the reduction of Coulomb energy due to TES will lead to an increase of total binding energy with respect to the $normal$ structure. The final consequence would be the decreasing of $V_{pn}(^{17}{\rm F)}$  (see Eq.~(\ref{eq:dv5})), giving a negative value of $\Delta V_{pn}$ as shown in Fig.~\ref{fig:DdVpn}.  


As mentioned above, the mirror symmetry of $V_{pn}$ is a prerequisite of the charge-symmetric Kelson-Garvey relation which is usually used for the mass prediction of proton-rich nuclei. The mirror-symmetry breaking of $V_{pn}$ implies that the predicted mass for a proton-unbound or proton-halo nucleus could be overestimated. 
Indeed, a significant deviation of predicted masses using the charge-symmetric Kelson-Garvey mass relations has been observed in Ref.~\cite{Comay1988} for some unstable proton-rich nuclei.  A large deviation in light proton-rich nuclei ($A<20$) also occurs in the mass predictions using the improved Kelson-Garvey mass relations ~\cite{Tian2013}. 

\section{Mirror-symmetry breaking of $V_{pn}$ investigated by Gamow shell model} \label{GSM}

\begin{figure*}[htbp]
	\centering
	\includegraphics[scale=0.6]{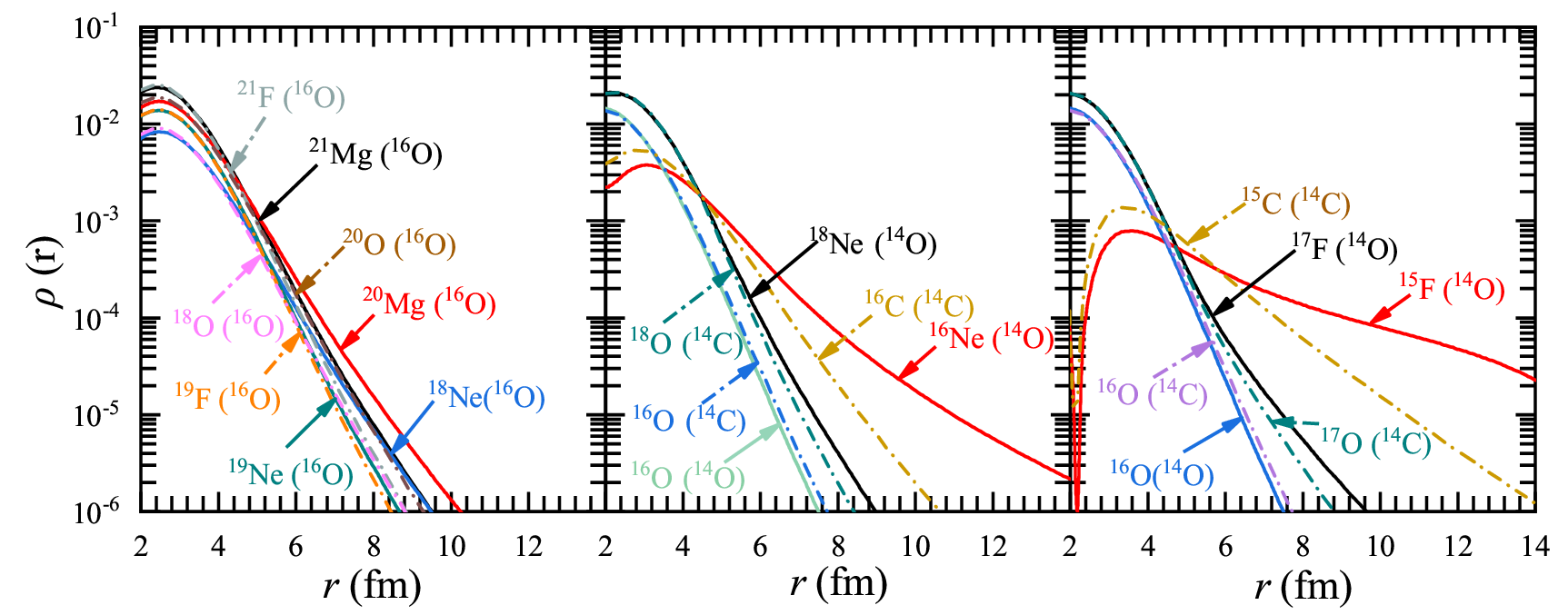}
	\caption{Density distribution calculated using Gamow shell model~\cite{PhysRevLett.89.042502,Michel_GSM_book,physics3040062,Xie2023}. The radial density distribution of valence protons is shown for the nuclei used in calculating the $V_{pn}$ values of proton-rich nuclei.  Conversely, the radial density distribution of valence neutrons is depicted for the nuclei used in calculating the $V_{pn}$ of neutron-rich nuclei. The nucleus specified in the bracket represents the inner core adopted in the Gamow shell model calculations}
	\label{fig:density}
\end{figure*}

To further understand the so-called mirror-symmetry breaking of $V_{pn}$, we have performed the Gamow shell model (GSM) calculations for a quantitative analysis of $\Delta V_{pn}$ focusing on the influence of TES. The GSM approach, traditionally conceptualized within the framework of a core plus valence particles, parallels the methodology of standard shell model calculations. The fundamental theoretical construction entering GSM is the one-body Berggren basis~\cite{BERGGREN1968265}, which consists of bound, resonance, and scattering single-particle (s.p.) states. However, only the bound s.p. states are used in the standard SM calculations. By generating a many-body basis from the comprehensive Berggren basis, we facilitate a nuanced representation of internucleon correlations through configuration mixing, thereby GSM calculations effectively address both the internucleon correlations and continuum coupling~\cite{PhysRevLett.89.042502,Michel_GSM_book,physics3040062,Xie2023}. 
Hence, GSM is a suitable approach to study weakly bound and unbound states. It has been successfully used to describe the one-neutron halos in $^{29,31}$Ne~\cite{LI2022137225}, the two-neutron halos in $^{6}$He~\cite{PhysRevC.84.051304}, $^{11}$Li~\cite{PhysRevC.101.031301} and $^{31}$F~\cite{PhysRevC.74.054305}, as well as the resonance structures of dripline nuclei~\cite{PhysRevC.106.L011301,PhysRevC.104.L061306,PhysRevC.104.024319,PhysRevC.103.034305,PhysRevC.100.054313,PhysRevLett.127.262502}, such as $^{7}$H  and $^{28}$O.
Here 
we employ GSM to probe the 
$p$-$n$ interactions in mirror nuclei and the mechanisms for mirror-symmetry breaking of $V_{pn}$ in the dripline nuclei, thus advancing our understanding of nuclear structure at the extremes of stability.

We take $^{21}$Mg, $^{17}$F, and $^{18}$Ne and their mirror partners as examples, since the $V_{pn}$ values of $^{21}$Mg-$^{21}$F pair exhibit small mirror asymmetry, and the ${V_{pn}}'$s of $^{17}$F-$^{17}$O and $^{18}$Ne-$^{18}$O pairs bear significant mirror-symmetry breaking. A two-body interaction in the chiral effective field theory is adopted in the real GSM calculations~\cite{PhysRevC.100.064303}. For the isospin non-conserving part of Hamiltonian, only the Coulomb force is considered, while the contribution of the isospin-dependent part of nuclear interaction to TES is small, which has been neglected in the present GSM calculations.
For the nuclei associated with the ${V_{pn}}'$s of $^{21}$Mg-$^{21}$F pair, GSM calculations are conducted with $^{16}$O as the inner core. 
For the nuclei associated with the ${V_{pn}}'$s of $^{17}$F and $^{18}$Ne, $^{14}$O is used as an inner core, while for their mirror nuclei, $^{14}$C inner core is used. The adopted Hamiltonian is optimized based on the experimental data.
The calculated $\Delta V_{pn}$ values are $-71$ keV for $^{21}$Mg, $-473$ keV for $^{17}$F, and $-222$ keV for $^{18}$Ne, which show good agreement with the experimental values of $-66$ keV, $-527$ keV, and $-200$ keV, respectively.

Our GSM calculations reveal significant mirror-symmetry breaking of $V_{pn}$ in $^{17}$F and $^{18}$Ne. To delve deeper into the mechanism behind this symmetry breaking, we analyzed the radial density distributions of nuclei involved in the $\Delta V_{pn}$ calculations for $^{21}$Mg, $^{17}$F and $^{18}$Ne. The results are depicted in Fig.~\ref{fig:density}.
For the nuclei associated with the ${V_{pn}}'$s of proton-rich nuclei, we present the radial density distribution of valence protons.  Conversely, the radial density distribution of valence neutrons is depicted for the nuclei associated with the ${V_{pn}}'$s of neutron-rich nuclei.
One can note that densities are complex for resonance states because these eigenstates have a complex energy. 
However, their imaginary part is small compared with their real part in our calculations, except for $^{15}$F, so we only consider real parts when dealing with densities in this work.
The calculated density decreases fast for all nuclei associated with $^{21}$Mg due to the bound character of those nuclei. 

As nuclei approach or beyond the dripline, they become weakly bound or unbound, leading to a more extended density distribution compared to bound nuclei. Specifically, the calculated density distributions for $^{15}$C and $^{15}$F are more extended than those of other nuclei associated with the $\Delta V_{pn}$ calculations for $^{17}$F.
Notably, $^{15}$C, characterized as a one-neutron halo, has its valence neutron occupying the weakly bound $\nu 1s_{1/2}$ orbital above the inner core of $^{14}$C. In contrast, $^{15}$F, an unbound nucleus, features its last proton in the unbound $\pi 1s_{1/2}$ orbital. Our calculations reveal that the radial density distribution of $^{15}$F is more extended than that of $^{15}$C.
Moreover, the GSM calculations indicate an oscillation in the asymptotic density distribution of $^{15}$F, a characteristic stemming from its unbound state.
A similar phenomenon is observed in the GSM calculations of the density of unbound oxygen dripline isotopes~\cite{PhysRevC.103.034305}. 

As delineated in Ref.~\cite{PhysRevC.100.064303}, the density distribution significantly influences the Coulomb contribution in nuclear systems. Specifically, a more extended density distribution reduces the Coulomb contribution, leading to a more bound nuclear system compared to the one with a localized density distribution. This observation aligns with the TES effect observed in the excited states of mirror nuclei, where the excitation energy of states in the proton-rich nucleus is typically lower than that of its neutron-rich counterparts, such as the $1_1^+$ state of $^{22}$Al and $^{22}$F mirror nuclei~\cite{Sun2024}. In summary, an unbound nucleus, characterized by an extended density distribution, inherently possesses more binding energy relative to a deeply bound nucleus. 

The results align with the notable mirror-symmetry breaking observed in $V_{pn}$ for $^{17}$F, which is characterized by a substantially negative $\Delta V_{pn}$, as discussed in Sec.~\ref{Breaking}. A comparable scenario is evident in the $\Delta V_{pn}$ for $^{18}$Ne, where $^{16}$Ne is unbound against 
two-proton decay. Our GSM calculations further reveal that the density distribution of $^{16}$Ne in the asymptotic region is more extended than that of its mirror nucleus, $^{16}$C.
Additionally, a large absolute value of $\Delta V_{pn}$ typically corresponds to scenarios where more than one proton-rich nucleus is unbound, while its mirror neutron-rich nucleus remains deeply bound, leading to significant disparities in their respective radial density distributions.

\section{Summary}\label{Summary}
In summary, thanks to the new mass values, especially those provided by B$\rho$-IMS, we systematically analyzed the effective proton-neutron interactions, $V_{pn}$, in the mirror nuclei along the $N=Z$ line. When all the nuclei involved are bound 
a mirror symmetry of $V_{pn}$ was revealed at a confidence level of $\sigma=32$ keV. 
This is expected in view of the charge symmetry and the charge independence of the nuclear force. The mirror symmetry of $V_{pn}$ supports the charge-symmetric Kelson-Garvey relations. 
However, for a few mirror pairs, such mirror symmetry is apparently broken as a proton-unbound nucleus is used in extracting the $V_{pn}$ value. 
The mirror symmetry breaking of $V_{pn}$ may induce deviations in the mass predictions using the charge-symmetric Kelson-Garvey mass relations.   

To understand mechanisms of the mirror-symmetry breaking of $V_{pn}$, 
we performed systematic Gamow shell model calculations for ${V_{pn}}'$s of the mirror nuclei. 
Good agreement of the $\Delta V_{pn}$ value between the calculation and experiment is intriguing. 
Our calculations confirmed that the significant mirror-symmetry breaking of $V_{pn}$ is mainly caused by the TES of a proton-unbound nucleus, in which the radial density distribution is more extended than that of deeply-bound neutron-rich mirror partner.  


\begin{acknowledgments}
	This work is supported in part by the National Key R$\&$D Program of China (Grant No. 2023YFA1606401), the Strategic Priority Research Program of Chinese Academy of Sciences (Grant No. XDB34000000), the Youth Innovation Promotion Association of the Chinese Academy of Sciences (Grants No. 2021419), the NSFC (Grants No. 12475128, No. 12305126, No. 12135017,  No. 12121005,  No. 12205340) and the CAS Project for Young Scientists in Basic Research (Grant No. YSBR-002). J. G. Li. is supported by the Gansu Natural Science Foundation under Grant No. 22JR5RA123. The numerical calculations were made on Hefei advanced computing center.
\end{acknowledgments}


\bibliography{3NF-ref.bib}

\end{document}